\documentclass[prl,aps,twocolumn,floatfix,superscriptaddress]{revtex4-1}
\usepackage{eso-pic,calc}
\usepackage{graphicx}
\usepackage{amsmath, amsthm, amssymb}
\usepackage{epsfig}
\usepackage{bm}
\usepackage{color}
\usepackage{bbding}
\usepackage{wasysym}

\usepackage[colorlinks=True]{hyperref}  
\hypersetup{
 	colorlinks=true,  
 	linkcolor=blue,  
 	citecolor=blue, 
 	filecolor=magenta,   
 	urlcolor=blue         
 }

\begin{document}
\title{$1/f$ noise in the Ising model}

\author{Rahul Chhimpa}
\affiliation{Department of Physics, Institute of Science,  Banaras Hindu University, Varanasi 221 005, India}

\author{Avinash Chand Yadav\footnote{jnu.avinash@gmail.com}}
\affiliation{Department of Physics, Institute of Science,  Banaras Hindu University, Varanasi 221 005, India}

\begin{abstract}
{We simulate the $N$-spin critical Ising model on a square lattice using Glauber dynamics and consider the typical one-unit time equal to $N$ single-spin-flip attempts. The divergence of correlation time with the linear extent of the system results in critical slowing down, a challenge to equilibration because the spin configurations generated in such a way are temporally correlated. We examine temporal correlations in the number of accepted spin flips and show a signature of non-trivial long-time correlation of a logarithmically decaying form or the corresponding power spectral density follows canonical $1/f$ noise.}
\end{abstract}
\maketitle

{\it Introduction}--
One of the puzzling features is understanding long-range temporal memory that emerges in processes as diverse as voltage fluctuations across an electronic device~\cite{PhysRev.26.71} to human cognition~\cite{doi:10.1126/science.7892611} or neuronal signals~\cite{ PhysRevLett.89.158101, PhysRevLett.97.118102, pettersen_2014, PhysRevLett.94.108103, Jensen_2021} to musical rhythms~\cite{levitin_2012}. Such signals display intriguing scaling of $1/f$ noise~\cite{RevModPhys.53.497} in their power spectrum. Efforts to uncover the underlying explanation have revealed the existence of a few routes. On the one hand, the superposition of several independent exponentially relaxing processes, even with uniformly distributed relaxation time~\cite{milotti20021}, or a simple nonlinear transformation of an exponentially relaxing process can yield $1/f$ noise~\cite{Yadav_2013, yadav_2017}. On the other hand, at the critical point of a second-order phase transition, the divergence of correlation length and time in the thermodynamic limit results in an emergent long-range space-time correlation~\cite{nishimori_2011}. Self-organized critical (SOC) systems dynamically evolve to reside at criticality~\cite{PhysRevLett.59.381, Christensen_2005} and thus can naturally explain the $1/f$ noise in many contexts, such as sandpile~\cite{PhysRevLett.83.2449, PhysRevE.85.061114, PhysRevLett.82.472}, fossil data~\cite{Ricard_1997}, and species evolution~\cite{PhysRevE.108.044109, PhysRevE.110.034130}, although SOC is not a necessary condition. 

The paradigmatic Ising model~\cite{Ising_1925} explains the ferromagnet to paramagnet (order-disorder) phase transition when the temperature crosses the critical point. Amazingly, the Ising model retains special attention in statistical physics for offering simple insight into generic critical phenomena~\cite{Krapivsky_2010} and has surprisingly diverse applications~\cite{PhysRevLett.87.265701, PhysRevE.66.016104,  PhysRevResearch.2.033182, PhysRevResearch.2.013199, PhysRevE.98.022138, Henkel_2008, Puri_2009}. The most fascinating aspect is the notion of universality class~\cite{RevModPhys.46.597, RevModPhys.70.653}, as different systems, irrespective of being microscopically different, can share the same set of scaling exponents at the critical point. The two-dimensional Ising model of magnets and the majority-vote model of opinion dynamics belong~\cite{PhysRevE.95.012101} to the same universality class. As is known, the universality feature depends on spatial dimensionality, the range of the interactions, and the system's symmetry.

\begin{figure}[t]
  \centering
       \scalebox{1.06}{\includegraphics{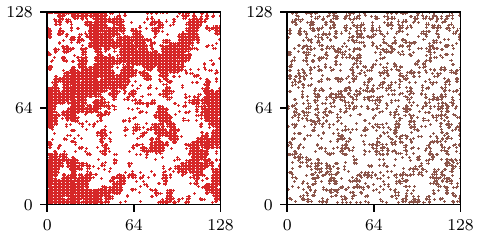}}
  \caption{Left panel: A typical spin configuration with $s_i = 1$ (in color) for the critical Ising model on a square lattice with $N = 2^{14}$. Right panel: The colored sites mark the spin-flip. }
  \label{fig_conf_1}
\end{figure}

We show numerical evidence of the intriguing $1/f$ noise with a spectral exponent of 1 or logarithmically decaying temporal correlation in the Ising model. Unexpectedly, the $1/f$ noise feature does not seem to depend on the detail of the underlying regular lattice structure. Numerically, we simulated the Ising model on a square lattice at the critical temperature using single-spin flip Glauber dynamics~\cite{glauber_1963}. The observable that we examine is the number of accepted spin flips  (cf. Fig.~\ref{fig_conf_1}) recorded as a function of time, where the time increases in the unit of $N$ proposed spin flips. Also, the number of accepted spin flips follows a Gaussian distribution with the system size scaling for the mean $\sim N$ and the variance $\sim N\ln{N}$.

The correlation time diverges for the critical Ising model~\cite{PhysRevLett.76.4548, nishimori_2011}, and one can naturally expect long-time correlation in various quantities, as previously well-studied for the total magnetization~\cite{liu_2023}.
We particularly emphasize that the emergence of logarithmic correlation for the number of accepted spin flips is strikingly intriguing, as criticality does not correspondingly imply a spectral exponent value of 1.  So far, such a simple observation seems overlooked, partly because the dynamic aspects remain challenging due to critical slowing down~\cite{herrmann_2021}, and the equilibrium properties of the Ising model are well-studied. Many other systems naturally exhibit logarithmic relaxation~\cite{doi:10.1073/pnas.1120147109}, for example, volume relaxation of crumpling paper~\cite{PhysRevLett.88.076101, PhysRevLett.130.258201} and frictional strength~\cite{Oded_2010}.

{\it The Ising model}--
We first recall the Ising model and its preliminary.
The model consists of $N$ spins arranged on a regular lattice. Each site can take either up spin $s_i = 1$ or down spin $s_i = -1$. 
The Hamiltonian for the classical Ising model is 
\begin{equation}
E = -J\sum_{\langle i,j\rangle}s_is_j -H\sum_i s_i,\nonumber
\end{equation}
where the first term represents the ferromagnetic interaction with a constant coupling strength $J>0$, and the second is the influence of an external magnetic field $H$. The symbol $\langle i,j\rangle$ denotes the nearest neighbors (nn) interaction.

The equilibrium properties can be deduced using canonical ensemble formalism~\cite{Christensen_2005}. The probability of a spin configuration of the model is $\exp(-E/k_BT)/Z$, where $k_B$ is the Boltzmann constant and $T$ is the temperature of the heat bath. The average magnetization per spin $m = \langle M \rangle/N$, where $M = \sum_i s_i$, serves as an order parameter. The symbol $\langle \cdot \rangle$ denotes the ensemble average. Physically relevant quantities (response functions) are the susceptibility $k_BT\chi = [\langle M^2 \rangle-\langle M \rangle^2]/N$ and the heat capacity. 
The one-dimensional (1D) Ising model remains solvable, although no phase transition occurs at a finite temperature. The model indeed shows a second-order phase transition for dimensions ${\rm D}\geq 2$, and the upper critical dimension is ${\rm D}_u = 4$ beyond which the mean-filed theory (MFT) is valid~\cite{PhysRevLett.28.240}. Interestingly, the 2D Ising model on a square lattice is also exactly solvable~\cite{onsager_1944}, yielding the critical temperature $k_BT_c/J = 2/\ln{[1+\sqrt{2}]}$.

In the vicinity of the critical point, the thermodynamic properties show scaling feature~\cite{nishimori_2011} as a function of reduced control parameters $\epsilon = (T-T_c)/T_c$ and $h = H/k_BT$ as $m(\epsilon<0, h\to 0) \sim |\epsilon|^{\beta}$ and $\chi \sim |\epsilon|^{-\gamma}$. The two-point spin-spin correlation function varies as $G^{(2)}(r) \sim 1/r^{2-{\rm D}+\eta}\exp(-r/\xi)$, where the correlation length diverges as $\xi \sim |\epsilon|^{-\nu}$. Only two critical exponents are independent since the exponents satisfy scaling relations~\cite{duminilcopin2022100yearscriticalising}.
The critical exponents are $\nu = 1$ and $\eta = 1/4$ in 2D and $\nu = 1/2$ and $\eta = 0$ for the MFT~\cite{nishimori_2011}.

\begin{figure}[t]
  \centering
       \scalebox{1}{\includegraphics{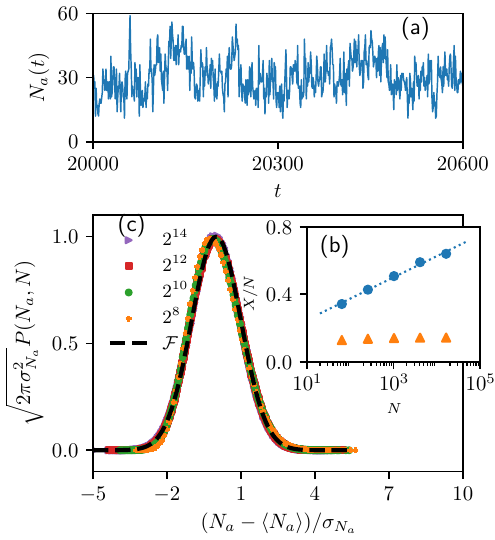}}
  \caption{For the critical Ising model on a square lattice with PBC: (a) A typical realization of the fluctuations in the number of accepted spin flips $N_a(t)$ with a system size $N = 2^8$. (b) The system size dependence of the mean ($\blacktriangle$) as $\langle N_a\rangle/N \sim {\rm constant}$ and the variance (\CIRCLE) as $\sigma_{N_a}^{2}/N \sim \ln{N}$. The dotted line represents the best-fit function $\sigma_{N_a}^{2}/N = 0.13(1)\ln{N}+0.12(2)$. (c) The data collapse of the probability distribution function, obtained by plotting $\sqrt{2\pi \sigma_{N_a}^{2} } P(N_a, N)$ with the shifted and scaled variable $(N_a - \langle N_a\rangle)/\sigma_{N_a}$, shows the normal distribution. The thick-dashed curve is $\mathcal{F}(u) = \exp(-u^2/2)$. The total number of events employed to determine the probability distribution is $2^{22}$.}
  \label{fig_pdf1}
\end{figure}

 {\it Simulation methods}-- 
The simple sampling of the Monte Carlo simulation method is not feasible since the number of micro-states grows exponentially $\sim 2^N$. The efficient algorithms based on importance sampling are ${\rm M(RT)^2}$~\cite{metropolis_1953} or Glauber dynamics~\cite{glauber_1963} (local dynamics) that satisfy the detailed balance condition~\cite{newmann_1999, binder_2000, gould_2007, kotze_2008}. In each Monte Carlo move, a randomly selected spin-flip $s_i\to -s_i$ changes energy as $\Delta E = 2Js_i\sum_{j\in {\rm nn}}s_j$. The acceptance of the spin-flip occurs with a transition probability  
\begin{equation}
W(s_i \to -s_i) = \begin{cases} {\rm min}\left\{1, \exp\left(-\frac{\Delta E}{k_BT}\right)\right\},~~{\rm M(RT)^2}, \\ {\left[1+\exp\left(\frac{\Delta E}{k_BT}\right)\right]}^{-1},~{\rm Glauber~method}. \end{cases}\nonumber
\end{equation}
Numerically generate a random number $x$ between 0 and 1 with uniform probability density. Accept the spin-flip if $x<W$; otherwise, reject the proposed spin-flip. The Monte Carlo step (MCS) or the time $t$ is defined in the unit of $N$ such proposed spin flips.  

At the critical point, the linear extent of the system size limits the correlation length in the model $\xi \sim |\epsilon|^{-\nu} \sim N^{1/{\rm D}}$, and the correlation time consequently varies as $\tau_0 \sim  |\epsilon|^{-\nu z} \sim \xi^z  \sim N^{z/{\rm D}}$, where $z$ is the dynamical exponent~\cite{PhysRevLett.76.4548}, with a mean-filed value $z = 2$~\cite{nishimori_2011} and $z \approx 2.16$ in 2D~\cite{PhysRevLett.76.4548, liu_2023}. The divergence of the correlation time implies that the equilibration time also diverges since the spin configurations generated via such local (single-spin flip) Monte Carlo method remain correlated in time.  The feature is referred to as critical slowing down.

\begin{figure}[t]
  \centering
       \scalebox{1.0}{\includegraphics{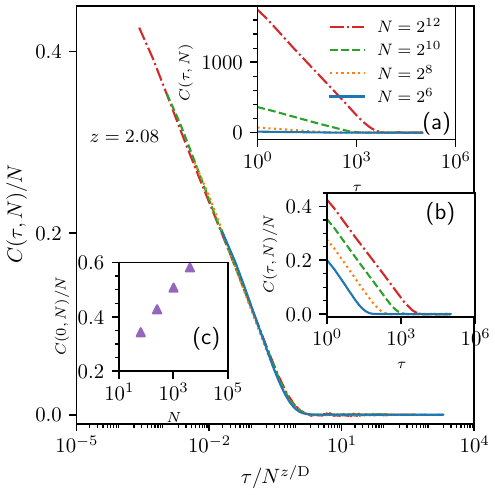}}
  \caption{(a) The two-time autocorrelation function $C(\tau, N)$ of $N_a(t)$ for different system sizes $N$ on a linear-logarithmic scale. Numerically, we computed it as $C(\tau, N) = \langle (\mathcal{T}-\tau)^{-1}\sum_{t=1}^{\mathcal{T}-\tau} N_a(t+\tau)N_a(t) \rangle$, with $\mathcal{T} = 2^{20}$. The number of samples used for the average is $10^2$, and the lag time $\tau$ runs from 1 to $10^5$. (b) The plot of scaled correlation function $C(\tau, N)/N$ with the lag $\tau$ implies the system size independence of the prefactor. (c) The scaled zero-lag correlation with the system size grows logarithmically. Main panel: The data collapse curve $\mathcal{G}(\tau/N^{z/{\rm D}})$ obtained by plotting $C(\tau, N)/N$ with $\tau/N^{z/{\rm D}}$ confirms a logarithmically decaying behavior [cf. Eq.~(\ref{eq_corr})]. }
  \label{fig_cr}
\end{figure}

\begin{figure}[t]
  \centering
       \scalebox{1}{\includegraphics{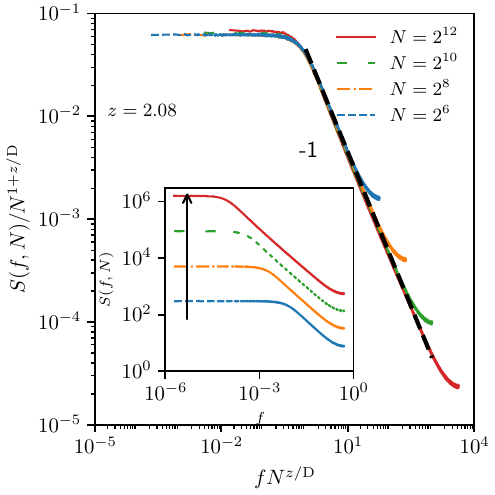}}
  \caption{Inset: The power spectra $S(f, N)$ for the number of accepted spin flips $N_a(t)$ with different system sizes $N$. The arrow marks the effect of increasing system size. We compute the power spectrum as $S(f, N) =  \lim_{\mathcal{T} \to \infty} \langle |\tilde{N_a} (f, N)|^2\rangle/\mathcal{T}$, where we implement the standard fast Fourier transformation method to determine the Fourier transformation $\tilde{N_a}(f, N)$. Each curve represents an average over $10^4$ independent realizations of the signal. Main panel: The scaled power spectra $S(f, N)/N^{1+z/{\rm D}}$ with reduced frequency $fN^{z/{\rm D}}$ for different system sizes collapse onto a single curve, yielding the scaling function $ \mathcal{H}(w)$ [cf. Eq.~(\ref{eq_psd})]. The thick-dashed straight line guides the slope of -1.}
  \label{fig_ps1}
\end{figure}

\begin{figure}[t]
  \centering
       \scalebox{1}{\includegraphics{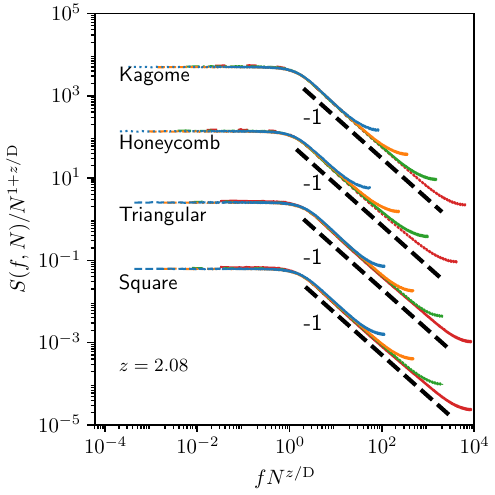}}
  \caption{Data collapse of the power spectra of $N_a(t)$ in the 2D Ising model on different lattices with helical boundary conditions. For clarity, we shift the curves vertically by a constant. The system sizes are $N = 2^5$, $2^7$, $2^9$, and $2^{11}$ in honeycomb and $N = 3\times \{2^4,  2^6, 2^8,  2^{10}\}$ in Kagom\'e lattices.}
  \label{fig_ps_top_1}
\end{figure}

\begin{figure}[t]
  \centering
       \scalebox{1}{\includegraphics{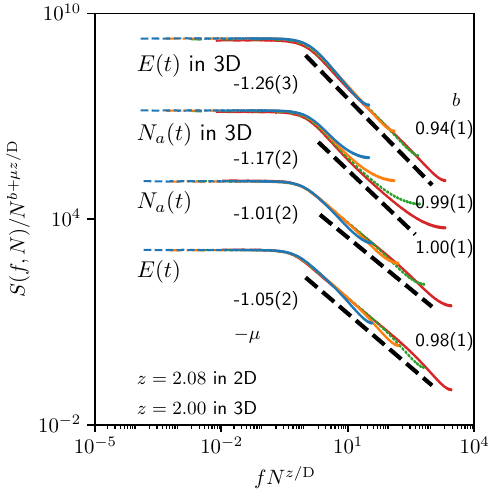}}
  \caption{The power spectra scaling functions for $E(t)$ and $N_a(t)$ in the 2D Ising model on a square lattice with PBC (the two bottom curves). For $N_a(t)$, the model is dynamically updated using ${\rm M(RT)^2}$ algorithm. We also show the scaling functions in 3D for $N_a(t)$ and $E(t)$, with system sizes $N = 2^6$, $2^9$, $2^{12}$, and $2^{15}$. The floating numbers are the respective critical exponents. The scaling functions reasonably agree with the straight lines within the statistical error.}
  \label{fig_ps_q_1}
\end{figure}

 {\it Results}-- 
In our simulations, we set $J/k_B = 1$, $H = 0$, and $T = T_c$. Starting with a random configuration, we apply the Glauber local update method to evolve the Ising model dynamically. 
Throughout our numerical results, we discard initial transients. First, we consider the Ising model on a square lattice with periodic boundary conditions (PBC). 
For a fixed system size $N$, the quantity we examine here is the number of accepted spin flips $N_a(t)$ as a function of time, measured in the unit of MCS.

Figure \ref{fig_pdf1}(a) depicts a typical temporal variation of the number of accepted spin flips $N_a(t)$ for a fixed system size $N$.
Numerically, we find a system size scaling for the mean and the variance as [cf. Fig.~\ref{fig_pdf1}(b)]
\begin{equation}
\langle N_a\rangle/N \sim {\rm constant}~~~{\rm and}~~~ \sigma_{N_a}^{2}/N \sim \ln{N}.
\end{equation}
The shifted and scaled variable $u = (N_a - \langle N_a\rangle)/\sigma_{N_a}$ follows the normal distribution $P(u) = (1/\sqrt{2\pi})\exp(-u^2/2)$ [cf. Fig.~\ref{fig_pdf1}(c)]. Also, the relative fluctuation varies as $\Delta_{N_{a}} \sim \sqrt{\ln{N}}/\sqrt{N}$.

As shown in Fig.~\ref{fig_cr}(a), the two-time autocorrelation function $C(\tau, N) = \langle N_a(t)N_a(t+\tau)\rangle$ for a fixed system size $N$ on a linear-logarithmic scale shows straight-line behavior below a correlation time and becomes zero beyond that, and when the system size increases, the correlation time and the slope of the straight line increase, implying that the correlation time and the prefactor both depend on the system size. The prefactor becomes independent of $N$ if we plot $C(\tau, N)/N$ with $\tau$, implying a logarithmically decaying behavior $C(\tau, N)/N \sim -\ln{|\tau|}$ below the correlation time [cf. Fig.~\ref{fig_cr}(b)]. Similarly, Fig.~\ref{fig_cr}(c) shows the behavior of zero-lag correlation or the variance that grows logarithmically with the system size as $C(\tau = 0, N)/N \sim \ln{N}$. It turns out, for $\tau \ll N^{z/{\rm D}}$,
\begin{equation}
\frac{C(\tau, N)}{N} = -A\ln{\left|\frac{\tau}{N^{z/{\rm D}}}\right|} = \mathcal{G}\left(\frac{\tau}{N^{z/{\rm D}}}\right), 
\label{eq_corr}
\end{equation}
where $A$ is a constant and the scaling function varies as $ \mathcal{G}(v) \sim -\ln{v}$ for $v\ll 1$ and 0 for $v \gg 1$.
Figure~\ref{fig_cr} (the main panel) shows a numerical verification of the data collapse curve $ \mathcal{G}(\tau/N^{z/{\rm D}})$ obtained by plotting the scaled correlation functions $C(\tau, N)/N$ with reduced time $\tau/N^{z/{\rm D}}$ for different system sizes.

As the process $N_a(t)$ is wide-sense stationary for finite system size, its power spectral density is related to the correlation function by Fourier transformation (the Wiener-Khinchin theorem) as $S(f, N) = \int C(\tau, N) \exp(-i2\pi f \tau) d\tau$ that yields
 \begin{equation}
S(f, N) = N^{1+z/{\rm D}} \mathcal{H}(fN^{z/{\rm D}}).
\label{eq_psd}
\end{equation} 
To determine the scaling function $ \mathcal{H}$, we plot scaled power spectra $S(f, N)/N^{1+z/{\rm D}}$ with reduced frequency $fN^{z/{\rm D}}$ and find that the scaling function $ \mathcal{H}(w)$ is constant for $w \ll 1$ and $\sim 1/w$ for $w \gg 1$ [cf. Fig.~\ref{fig_ps1}]. In turn, the power spectral density varies as 
 \begin{equation}
 S(f, N)/N \sim 1/f,~~~~{\rm for}~~~~1/N^{z/{\rm D}} \ll f \ll 1/2.
  \end{equation}

To uncover the extent of the $1/f$ noise feature for the process $N_a(t)$, we also examine the 2D Ising model on different regular lattices such as square, triangular ($T_c = 4/\ln 3$), honeycomb ($T_c = 2/\ln{[2+\sqrt{3}]}$)~\cite{Christensen_2005}, and Kagom\'e ($T_c = 2.143$)~\cite{KASSANOGLY2023170568} with a helical boundary condition~\cite{newmann_1999}. Although the critical temperature is a function of the underlying topology, the exponents do not change, reflecting the same universality class. As shown in Fig.~\ref{fig_ps_top_1}, we find the $1/f$ noise holds for the noisy process $N_a(t)$ in the 2D Ising model for different regular lattices. 

For the Ising model on a square lattice with PBC, we also present in Fig.~\ref{fig_ps_q_1} a comparison of the underlying temporal correlations for different noisy processes such as $E(t)$ and $N_a(t)$. We use the ${\rm M(RT)^2}$ local dynamic for $N_a(t)$. We find that different signals show the presence of long-time correlations; however, the spectral exponent is very close to 1 in the case of $N_a(t)$.

Finally, we examine the effect of dimensionality for the $1/f$ noise in the signals $N_a(t)$ and $E(t)$. Using the Glauber dynamics, we simulate the 3D Ising model ($T_c = 4.51152$)~\cite{Christensen_2005} on a cubic lattice with helical boundary conditions~\cite{newmann_1999}. The $1/f$ noise in the signals also holds, as numerically verified in dimensions 2 and 3, but the spectral exponents differ marginally. 
Intuitively, Eq.~(\ref{eq_psd}) assumes a general form as 
$S(f, N) = N^{b+\mu z/{\rm D}} \mathcal{H}(fN^{z/{\rm D}})$,
with $\mathcal{H}(w)$ constant for $w\ll 1$ and $\sim w^{-\mu}$ for $w\gg 1$. Here $\mu$ is the spectral exponent. The power typically scales as $\sim N^b$ above the cutoff frequency with $b \approx 1$ [cf. Eq.~(\ref{eq_psd})].
Moreover, the system size scaling of the power in the low-frequency regime and the estimate of $b$ provide a numerical value of the spectral exponent for a given dynamical exponent (cf. Fig.~\ref{fig_ps_q_1}).

 {\it Conclusions}-- 
To conclude, we employed the single spin-flip Glauber algorithm to simulate the critical Ising model on a square lattice. Remarkably, the two-time autocorrelation function for the number of accepted spin-flips decays logarithmically [cf. Eq.~(\ref{eq_corr})], and the process then exhibits the canonical $1/f$ noise [cf. Eq.~(\ref{eq_psd})]. Although the critical slowing down in the model is a challenge to equilibration~\cite{newmann_1999}, this aspect offers a fresh opportunity as a natural illustration of the canonical $1/f$ noise.

In extension, the $1/f$ noise is robust for different regular lattices in 2D, including triangular, honeycomb, and Kagom\'e. Beyond this, the $1/f$ noise with a spectral exponent value of 1 does not seem hyper-universal, as verified in dimensions 2 and 3. It would be interesting to examine the extent of the observed $1/f$ noise in related externally tuned critical dynamics. An analytical prediction of the spectral exponent remains a challenging and pertinent question for future studies. 

{\it Acknowledgments}--
RC acknowledges UGC, India, for financial support through a Senior Research Fellowship. We also acknowledge the Supercomputing Centre, IIT BHU, Varanasi, for supporting the computer simulations on the PARAM Shivay high-performance supercomputer.

\bibliography{s1sources}
\bibliographystyle{apsrev4-1}

\end{document}